\documentclass[showpacs,showkeys,amsmath,amssymb,preprint,nofootinbib,prd,superscriptaddress]{revtex4}
\usepackage{graphicx}
\usepackage{epsfig}
\usepackage{bm}
\usepackage{float}
\usepackage{fontenc}
\usepackage{graphicx}
\usepackage{bm}
\usepackage{amsmath}
\def\beq{\begin{eqnarray}}

\def\eeq{\end{eqnarray}}

\begin{document}

\title[perturbation]{Hairy black hole stability under odd parity perturbations in the Einstein-Gauss-Bonnet model}

\author{Gilberto Aguilar-P\'erez\footnote{E-mail: 218570370@alumnos.fcfm.buap.mx}}
\affiliation{Facultad de Ciencias F\'\i sico-Matem\'aticas, Benem\'erita Universidad Aut\'onoma de Puebla, Puebla, M\'exico}

\author{Miguel Cruz\footnote{E-mail: miguelcruz02@uv.mx}}
\affiliation{Facultad de F\'\i sica, Universidad Veracruzana 91000, Xalapa, Veracruz, M\'exico}

\author{Samuel Lepe\footnote{E-mail: samuel.lepe@pucv.cl}}
\affiliation{Instituto de F\'\i sica, Facultad de Ciencias, Pontificia Universidad Cat\'olica de Valpara\'\i so, Av. Brasil 2950, Valpara\'\i so, Chile}

\author{Israel Moran-Rivera\footnote{E-mail: israelrivera465@gmail.com}}
\affiliation{Facultad de F\'\i sica, Universidad Veracruzana 91000, Xalapa, Veracruz, M\'exico}

\date{\today}

\begin{abstract}
Supported by the use of a regular scalar field we find a black hole solution in the Einstein-Gauss-Bonnet model. From the obtained solution we can recover the Schwarzschild black hole as in other works. Later, by implementing the odd parity perturbations method we study the stability of the linearized equations of motion of the model, we find the explicit form of the Regge-Wheeler potential and we explore the condition of vanishing perturbations at the horizon of the black hole. We test the stability of the obtained solution by checking the positivity condition of the Regge-Wheeler potential. Finally, we show that the stability of model depends on the value of a parameter introduced in the profile for the scalar field.    
\end{abstract}



\maketitle
\section{Introduction}
\label{sec:intro}

The inclusion of a scalar field in general relativity is the simplest modification that can be implemented to gravity and the resulting scheme it is usually known as scalar-tensor gravity; given that the scalar field is an elementary field in nature, over the years the scalar-tensor models have been subject of great acceptance and applicability in several branches of physics, being cosmology the most outstanding. For an introductory review on the role of scalar fields in cosmology the Ref. \cite{faraoni} can be useful. For instance, the current homogeneity observed in the universe can be explained in simple manner by considering the existence (before matter domination) of a primordial scalar field that at early times drives during a very short period an accelerated superluminal expansion, this scenario is commonly named inflation model \cite{guth}.\\ 
At this point it is natural to question about the origin of scalar fields, the answer is still unclear, however it seems that a geometrical origin for the scalar field can be established when higher order invariants are considered in the gravitational action, one example of this situation is given by the $f(R)$ gravity, in which can be found an equivalence between this and the Brans-Dicke model \cite{soti}. Another example of the aforementioned situation is given in the context of brane models \cite{cruz1}, for a certain type of Born-Infeld action the bending modes of the brane are characterized by a scalar field that can be related to the Galileon models, i.e., scalar fields with second-order dynamics \cite{galileon1, galileon2, galileon3, galileon4, galileon5}. Then those models including higher order invariants or derivatives represent an interesting scenario to test gravity. The Horndeski gravity \cite{horn} is one example of this kind of model and represents the most general scalar-tensor theory with second-order equations of motion, as in general relativity. This latter characteristic is also preserved in the Lovelock gravity \cite{love}. It is worthy to mention that despite the presence of high-order invariants in the last two examples, only by the fact that preserve second-order equations of motion some instabilities can be avoided such as the Ostrogradski ghost.\\

An interesting model results from the coupling of the scalar field and the Gauss-Bonnet invariant that emerges in the Lovelock gravity. This geometrical invariant has cuadratic terms of the curvature and it is well established that the term by itself is not dynamic in four dimensions, i.e. it is topological, however when coupled to the scalar field becomes dynamical and governed by second-order equations of motion, as we will see later. This coupling have been studied widely at cosmological level, see for instance the Refs. \cite{rad1, rad2, mota, chakra, odint}.\\ 

On the other hand, one peculiarity of this model was found lately, contrary to what usually happens in scalar-tensor theories, black hole or neutron stars solutions in this model could be distinguishable from those obtained in general relativity and this is due to the {\it scalarization} effect. The no-hair theorems implies that standard scalar-tensor theories will have the same stationary and asymptotically flat black hole solutions as general relativity \cite{thomas}, however recently has been found that models with couplings between the scalar field and some curvature invariants can induce an effective mass on the scalar mode at perturbative level \cite{doneva, sotiriou, tps}, then objects of general relativity such as neutron stars could present a scalar hair \cite{neutron}. In consequence, in Ref. \cite{babichev} within the context of the Gauss-Bonnet coupling it is stated that the scalarization effect could be catastrophic for the stability of compact objects obtained in general relativity, since the mass of the scalar mode can be tachyonic. Therefore, becomes relevant to study the perturbative effects on solutions obtained in this model, some black hole solutions can be found in Refs. \cite{bh1, bh2, bh3}.\\ 

In this work we will approach the stability problem for the Gauss-Bonnet coupling with the use of a standard method known as odd parity perturbations, see for instance the Refs. \cite{odd, anabalon, cisterna, manuel, gan}. This analysis will be carried out over a black hole type solution obtained in this work, where the use of a profile for the scalar field which principal characteristic is that is regular everywhere was considered.\\        

This paper it is constituted by four sections. In Section \ref{sec:hbh} we give some generalities of the model, we write the equations of motion that govern the dynamics and introduce the profile for the scalar field, as well as the form of the metric with which the generalized Einstein equations are solved. We discuss some particularities of the obtained solution, given in terms of the hypergeometric function and we show that the Schwarzschild black hole can be recovered from this. In Section \ref{sec:egb} we implement the method of odd parity perturbations for the Einstein-Gauss-Bonnet model and we explore the stability modes with the use of the positivity criterion for the potential obtained in the generalized Regge-Wheeler equation. We show that the perturbations vanish at the horizon of the black hole and by the election of some values for the parameters involved in the model we illustrate the shape the potential. Finally, in Section \ref{sec:final} we give the final comments of our work.  

\section{Hairy black hole solution}
\label{sec:hbh}

In this work we will consider the action
\begin{equation}
S\left[g_{\mu \nu}, \phi \right]= \int d^{4}x \sqrt{-g}\left[\frac{R}{2}-\frac{\left(\nabla\phi\right)^2}{2}+\dfrac{1}{8}f(\phi)R^2_{GB}-V(\phi)\right], 
\label{eq:action}
\end{equation}
where $g$ is the determinant of the metric tensor $g_{\mu \nu}$, $\left(\nabla\phi\right)^2 := g^{\mu\nu}\partial_{\mu}\phi\partial_{\nu}\phi$, is the kinetic term of the scalar field, $\phi$, $V(\phi)$ is the scalar field potential and $f(\phi)$ is an arbitrary function that mediates the non minimal coupling between the scalar field and the Gauss-Bonnet term, which is given as $R^{2}_{GB} = R^{2}-4R_{\mu \nu}R^{\mu \nu} + R_{\alpha \beta \gamma \delta}R^{\alpha \beta \gamma \delta}$. By performing a variation of action (\ref{eq:action}) with respect to the metric and the scalar field we obtain, respectively
\begin{align}
& E_{\mu \nu}:= G_{\mu\nu}+P_{\mu\alpha\nu\beta} \nabla^{\alpha\beta}f(\phi)-\partial_{\mu}\phi\partial_{\nu}\phi+g_{\mu\nu}\left[\dfrac{1}{2}(\partial\phi)^2+V(\phi)\right]=0, \label{eq:metric} \\
& \dfrac{1}{\sqrt{-g}}\nabla_{\mu}\left(\sqrt{-g}\nabla^{\mu}\phi\right)+\dfrac{1}{8}f'R^2_{GB} - V'(\phi)=0, \label{eq:kg}
\end{align}
being $G_{\mu \nu}$ the Einstein tensor and the prime stands for a derivative with respect to the scalar field, for simplicity in the notation we have the following definition
\begin{equation}
P_{\mu\alpha\nu\beta}=R_{\mu\alpha\nu\beta}+2g_{\mu[\beta}R_{\nu]\alpha}+2g_{\alpha[\nu}R_{\beta]\mu}+Rg_{\mu[\nu}g_{\beta]\alpha},
\end{equation}
the square brackets denote anti symmetry. From now on in this work, for simplicity we will consider the function, $f(\phi) = \phi$. Following the line of reasoning of Refs. \cite{joel, joel1}, in this section we find exact hairy black hole solutions considering a scalar field that is regular everywhere given by the following profile 
\begin{equation}
\phi(r)=\dfrac{1}{\sqrt{2}}\ln\left(1+\dfrac{\nu}{r}\right),
\label{eq:scalar}
\end{equation}
being $\nu$ a constant parameter, for the metric Ansatz we will consider
\begin{equation}
ds^2=-f(r)dt^2+\dfrac{dr^2}{f(r)}+r^{2}d\Omega^{2},
\label{eq:ansmetric}
\end{equation}
with coordinates $(t,r,z,\psi)$ and being $d\Omega^{2}$ the metric of the two dimensional spatial section with curvature parameter, $k$, which can take the values $0, \pm 1$. From the previous input, the equation of motion (\ref{eq:metric}) can be simplified to the following expressions
\begin{equation}
f(r) = \frac{r^{2}+\sqrt{2}k\left(1+\frac{2r}{\nu} \right)}{\sqrt{2}\left(1+\frac{2r}{\nu} \right)},
\label{eq:func}
\end{equation}
which is simply an algebraic expression and
\begin{equation}
\left(1-\frac{1}{r}\right)f''(r)-\frac{2\nu}{r^{3}}\left(2+\frac{1}{r}\right)f(r)+\frac{2\sqrt{2}\nu}{r^{2}}\left(\frac{1}{2}+\frac{k}{\sqrt{2}r^{2}}+\frac{\sqrt{2}k}{\nu r} \right)=0,
\label{eq:diff}
\end{equation}
where the prime denotes derivative with respect to the radial coordinate, $r$. The Eq. (\ref{eq:func}) can be obtained from $E^{t}_{t}-E^{r}_{r}$ and the Eq. (\ref{eq:diff}) from the components $E^{z}_{z}-E^{r}_{r}$ together with $E^{t}_{t}-E^{z}_{z}$. In this case we have $E^{z}_{z} = E^{\psi}_{\psi}$. Note that once we have an explicit expression for $f(r)$, the form of the scalar field potential can be obtained straightforwardly from the components $E^{z}_{z}$ or $E^{\psi}_{\psi}$, as a function of the radial coordinate. By considering the function $f(r)$ given in (\ref{eq:func}), the Eq. (\ref{eq:diff}) takes the form 
\begin{equation}
\left(1-\frac{1}{r}\right)f''(r)+\frac{4}{r^{3}}\left(1-\nu\right)f(r)=0.
\label{eq:diff2}
\end{equation}
Taking the change of variable $z := 1-r^{-1}$, the previous differential equation can be rewritten as follows
\begin{equation}
z(1-z)^{4}\frac{d^{2}f(z)}{dz^{2}}-2z(1-z)^{3}\frac{df(z)}{dz}+4(1-\nu)(1-z)^{3}f(z) = 0.
\end{equation}
With the change, $f(z) = z^{\alpha}(1-z)^{\beta}F(z)$ in the previous equation one gets
\begin{equation}
z(1-z)\frac{d^{2}F(z)}{dz^{2}} + \left[c-\left(a+b+1\right)z\right]\frac{dF(z)}{dz}-abF(z) = 0,
\label{eq:hyper}
\end{equation}
which corresponds to the hypergeometric differential equation for $F(z)$, in this case the following conditions must be satisfied
\begin{align}
& c = 2\alpha, \label{eq:integer} \\
& a+b = 2\alpha + 2\beta + 1,\\
& ab = \alpha(\alpha-1) + \beta(\beta-1) + 2\alpha + 2\beta + 2\alpha \beta - 4(1-\nu),
\end{align}
which gives
\begin{equation}
a = \frac{1}{2}\left(2\alpha + 2\beta \mp \sqrt{17-16\nu}+1 \right), \ \ \ \mbox{and} \ \ \ b = \frac{1}{2}\left(2\alpha + 2\beta \pm \sqrt{17-16\nu}+1 \right),
\label{eq:coeff}   
\end{equation}
and the exponents are given by $\beta = 0, -1$ and $\alpha = 0,1$. Note that the solutions for $a$ and $b$ impose the condition, $\nu \leq 17/16 \approx 1.06$. The general solution of Eq. (\ref{eq:hyper}) can be written as follows \cite{abramowitz}
\begin{equation}
F(z) = AF_{1}(a,b,c;z)+Bz^{1-c}F_{1}(a-c+1,b-c+1,2-c;z),
\label{eq:solution}
\end{equation}
where $A$ and $B$ are arbitrary constants and $F_{1}(a,b,c;z)$ is the hypergeometric function, then the function $f(z)$ takes the form
\begin{equation}
f(z) = Az^{\alpha}(1-z)^{\beta}F_{1}(a,b,c;z)+Bz^{1-c+\alpha}(1-z)^{\beta}F_{1}(a-c+1,b-c+1,2-c;z).
\label{eq:solution1}
\end{equation}
It is worthy to mention that for the change of variable considered we have $z(r \rightarrow 0)\rightarrow -\infty$, $z=0$ for $r=r_{H}=1$ and $z(r \rightarrow \infty) \rightarrow 1$. Note that we will have several cases once we consider the values of the exponents $\alpha$ and $\beta$ and the signs in the expressions for $a$ and $b$. As can be seen from the condition (\ref{eq:integer}), $c$ will be an integer with values $0$ or $2$, therefore the solution of the hypergeometric equation given in (\ref{eq:solution}) changes slightly. For $n$ being an integer number we have the following conditions: if $c = n \geq 2$ the second term of (\ref{eq:solution}) can not represent a solution of the hypergeometric equation and if $c = - n \leq 0$, the first term of (\ref{eq:solution}) can not be considered as solution of the hypergeometric equation \cite{seaborn}. For instance, if we consider the case $c = n \geq 2$, we can write
\begin{equation}
f(z) = Az(1-z)^{\beta}F_{1}(a,b,2;z) = Az(1-z)^{\beta}\sum^{\infty}_{n=0}\frac{(a)_{n}(b)_{n}}{(2)_{n}}\frac{z^{n}}{n!}, 
\label{eq:solution2}
\end{equation} 
where $(a)_{n} = a(a+1)(a+2)...(a+n-1)$ is the Pochhammer symbol, by considering $\beta = 0$ and truncating the previous sum at the first term, one gets
\begin{equation}
f(r) \approx A\left(1-\frac{1}{r}\right),
\end{equation} 
which corresponds to the Schwarzschild black hole with horizon event located at $r=1$, besides we have used the property $(a)_{0} =1$ and the change of variable defined previously. Something similar occurs in Ref. \cite{bh3}, the Schwarzschild black hole is recovered in one branch of the solution obtained in the discussion done by the authors. On the other hand, for $\beta = - 1$ the function $f(r)$ has the following behavior for the nth order term: $r(1-1/r)^{n+1}$. 

\section{Odd parity perturbations in the Einstein-Gauss-Bonnet model}
\label{sec:egb}

In order to study the odd parity perturbations, we adopt the following perturbed metric
\begin{equation}
ds^2=-A(r)dt^2+B(r)dr^2+C(r)\left[\dfrac{dz^2}{(1-k z^2)}+(1-k z^2)(d\varphi+k_1dt+k_2dr+k_3dz)^2\right].
\label{eq:pert}
\end{equation}
In general, $k_{1}, k_{2}$ and $k_{3}$ are functions of the coordinates $(t,r,z)$. On the other hand, $A(r)$, $B(r)$, $C(r)$ are some metric functions parameterizing the most general static background solution of a scalar-tensor theory. For the scalar field we will consider the following form
\begin{equation}
\phi=\phi_0(r)+\varepsilon\Phi(t,r,z), 
\label{eq:field}
\end{equation}
being $\phi_0$ the background field. Taking into account the metric (\ref{eq:pert}) and the scalar field (\ref{eq:field}), we can write the equations of motion (\ref{eq:metric}) at first order in $\varepsilon$ as follows
\begin{align}
& E_{r}^{t}=\dfrac{\varepsilon}{A}\left[\dfrac{k\partial_{r}}{C}+\phi_{0, r}-\dfrac{C_{r}^2\partial_{r}}{4BC^2}-\dfrac{A_{r}k}{2AC}+\dfrac{C_{r}^2A_{r}}{8ABC^2}\right]\partial_{t}\Phi+\mathcal{O}(\varepsilon^2)=0, \label{eq:eom1} \\
& E_{z}^{r}=-\dfrac{\varepsilon}{B}\left[\phi_{0 ,r}-\dfrac{C_{r}A_{r}\partial_{r}}{4BAC}+\dfrac{C_{r}^2A_{r}}{8BAC^2}\right]\partial_{z}\Phi+\mathcal{O}(\varepsilon^2)=0. \label{eq:eom2}
\end{align}
Note that the previous equations lead us to the conditions $\partial_{t}\Phi = \partial_{z}\Phi =0$. If we  consider the operation $E_{t}^{t}-E_{r}^{r}$, one gets
\begin{align}
E_{t}^{t}-E_{r}^{r} &= \dfrac{\varepsilon}{B}\left[-\dfrac{1}{2}\dfrac{kB_{r}}{BC}+\dfrac{C_{r}^3}{4BC^3} +2\phi_{0,r}+\dfrac{3}{8}\dfrac{C_{r}^2B_{r}}{B^2C^2}-\dfrac{1}{4}\dfrac{C_{r}^2\partial_{r}}{BC^2}+\dfrac{k\partial_{r}}{C} -\dfrac{1}{2}\dfrac{C_{r}C_{r,r}}{BC^2}+\dfrac{3}{8}\dfrac{C_{r}^2A_{r}}{BAC^2}\right. \nonumber \\
& \left. -\dfrac{1}{2}\dfrac{kA_{r}}{AC}\right]\partial_{r}\Phi+\mathcal{O}(\varepsilon^2)=0, 
\label{eq:eom3}
\end{align} 
therefore $\partial_{r}\Phi = 0$. Using these results we can simplify the form of the equation for $E_{r}^{r}$, one gets
\begin{equation}
E_{r}^{r}=\varepsilon V_{1}\Phi+\mathcal{O}(\varepsilon^2)=0,
\end{equation} 
which implies $\Phi = 0$, in previous equation we have defined 
\begin{equation}
V_{1}=\left. \dfrac{dV}{d\phi} \right|_{\phi=\phi_{0}}. 
\label{eq:eom4}
\end{equation}
It is possible to check that the remaining equations are satisfied up to linear order in $\varepsilon$ always that the following system is fulfilled
\begin{eqnarray}
E^{r}_{\varphi} &=& \partial_{z}\left[\dfrac{A}{C}\left(1-\dfrac{\phi_{0, r}A_{r}}{2AB}\right)\left(1-kz^2\right)^2\left(\partial_{z}k_{2}-\partial_{r}k_{3}\right)\right]\nonumber \\ 
&+& \partial_{t}\left[\left(1-\dfrac{\phi_{0,r}C_{r}}{2BC}\right)\left(1-kz^2\right)\left(\partial_{r}k_{1}-\partial_{t}k_{2}\right)\right]=0, \label{eq:eom5}\\
E_{\varphi}^{t} &=& \partial_{z}\left[C\sqrt{\frac{B}{A}}\left(1-kz^2\right)^2\left(1-\frac{\phi_{0r,r}}{B}+\dfrac{\phi_{0,r}B_{r}}{2B^2}\right)\left(\partial_{z}k_{1}-\partial_{t}k_{3}\right) \right]\nonumber \\
&+& \partial_{r}\left[\frac{C^{2}}{\sqrt{AB}}\left(1-kz^2\right)\left(1-\frac{\phi_{0,r}C_{r}}{2BC}\right)\left(\partial_{r}k_{1}-\partial_{t}k_{2}\right) \right]=0 \label{eq:eom6},\\
E^{z}_{\varphi} &=& \partial_{r}\left[C\sqrt{\dfrac{A}{B}}\left(1-\dfrac{\phi_{0,r}A_{r}}{2AB}\right)\left(\partial_{z}k_{2}-\partial_{r}k_{3}\right)\right]\nonumber \\ 
&+& \partial_{t}\left[C\sqrt{\dfrac{B}{A}}\left(1-\dfrac{\phi_{0r,r}}{B}+\dfrac{\phi_{0,r}B_{r}}{2B^2}\right)\left(\partial_{t}k_{3}-\partial_{z}k_{1}\right)\right]=0. \label{eq:eom7}
\end{eqnarray}
If we propose the following variable
\begin{equation}
\mathcal{Q}(t, r, z)=C\sqrt{\dfrac{A}{B}}\mathcal{P}(r)\left(1-kz^2\right)^2\left(\partial_{z}k_{2}-\partial_{r}k_{3}\right),
\end{equation}
where
\begin{equation}
\mathcal{P}(r)=\left(1-\dfrac{\phi_{0,r}A_{r}}{2AB}\right), \label{eq:var}
\end{equation}
the set of equations (\ref{eq:eom5})-(\ref{eq:eom7}) takes the form
\begin{align}
& \dfrac{\sqrt{AB}}{C^2\left(1-kz^2\right)\mathcal{W}}\dfrac{\partial\mathcal{Q}}{\partial z}=-\partial_{t}\partial_{r}k_{1}+\partial_{t}^2k_{2}, \label{eq:eomvar1}\\
& \dfrac{\sqrt{A}}{C\sqrt{B}\left(1-kz^2\right)^2\mathcal{S}}\dfrac{\partial\mathcal{Q}}{\partial r}=-\partial_{t}^2k_{3}+\partial_{t}\partial_{z}k_{1}. \label{eq:eomvar2}
\end{align}
For simplicity in the notation we have defined the following functions 
\begin{align}
& \mathcal{W}(r)=\left(1-\dfrac{\phi_{0,r}C_{r}}{2BC}\right),\label{eq:f1}\\
& \mathcal{S}(r)=\left(1-\dfrac{\phi_{0r,r}}{B}+\dfrac{\phi_{0,r}B_{r}}{2B^2}\right). \label{eq:f2}
\end{align}
Taking into account the combination $\partial_r(\ref{eq:eomvar2})+\partial_z(\ref{eq:eomvar1})$, one gets an equation in terms of the $\mathcal{Q}$ variable as follows
\begin{equation}
\dfrac{C^2\mathcal{W}}{\sqrt{AB}}\dfrac{\partial}{\partial r}\left[\dfrac{1}{C\mathcal{S}}\sqrt{\dfrac{A}{B}}\dfrac{\partial\mathcal{Q}}{\partial r}\right]+\left(1-kz^2\right)^2\dfrac{\partial}{\partial z}\left[\dfrac{1}{\left(1-kz^2\right)}\dfrac{\partial\mathcal{Q}}{\partial z}\right] = \dfrac{C}{A}\dfrac{\partial_{t}^2\mathcal{Q}}{\mathcal{P}}.\label{eq:forQ}
\end{equation}
In order to solve the previous equation, we consider the method of separation of variables, i.e., we consider $\mathcal{Q}=Q(r,t)D(z)$, yielding
\begin{align}
& \dfrac{C^2\mathcal{W}}{\sqrt{AB}}\dfrac{\partial}{\partial r}\left[\dfrac{1}{C\mathcal{S}}\sqrt{\dfrac{A}{B}}\dfrac{\partial Q}{\partial r}\right]-\lambda Q=\dfrac{C}{A}\dfrac{\partial_{t}^2Q}{\mathcal{P}}, \label{eq:sepa1} \\
& \left(1-kz^2\right)^2\dfrac{\partial}{\partial z}\left[\dfrac{1}{\left(1-kz^2\right)}\dfrac{\partial D}{\partial z}\right]=-\lambda D, \label{eq:sepa2}
\end{align}
where $\lambda$ is a constant which results from the method of separation of variables. For the case $k=1$ and considering $z=\cos\theta$, therefore the Eq. (\ref{eq:sepa2}) has a solution given in terms of the Gegenbauer polynomials as follows, $D(z)=C_{l+2}^{-3/2}\left(\theta\right)$, where $\lambda=\left(l-1\right)\left(l+2\right)$, with the condition $l\geq 1$. The relation between the Gegenbauer and Legendre polynomials is given as
\begin{equation}
C_{l+2}^{-3/2}\left(\theta \right)=\sin^3 \theta \dfrac{d}{d\theta}\dfrac{1}{\sin \theta}\dfrac{dP_{l}\left(\theta \right)}{d\theta}.
\end{equation}
If we consider the master variable $\Psi(r^*,t)=\left[C(r)\mathcal{S}(r)\right]^{-1/2}Q(r,t)$, being $r^{*}$ the Regge-Wheeler coordinate and $\dfrac{\partial}{\partial r}=\sqrt{\dfrac{B}{A}}\dfrac{\partial}{\partial r^{*}}$, the master equation (\ref{eq:sepa1}) takes the form
\begin{eqnarray}
& \dfrac{\partial^2\Psi}{\partial r^{*2}}+\left[\dfrac{1}{2\mathcal{S}}\dfrac{\partial^2\mathcal{S}}{\partial r^{*2}}-\dfrac{3}{4\mathcal{S}^2}\left(\dfrac{\partial\mathcal{S}}{\partial r^{*}}\right)^2-\dfrac{1}{2C\mathcal{S}}\dfrac{\partial\mathcal{S}}{\partial r^{*}}\dfrac{\partial C}{\partial r^{*}}+\dfrac{1}{2C}\dfrac{\partial^2C}{\partial r^{*2}}\right. \nonumber \\
& \left. - \dfrac{3}{4C^{2}}\left(\dfrac{\partial C}{\partial r^{*}}\right)^2-\lambda\dfrac{A\mathcal{S}}{C\mathcal{W}}\right]\Psi =\dfrac{\mathcal{S}}{\mathcal{P}\mathcal{W}}\partial^2_{t}\Psi.
\label{eq:master}
\end{eqnarray}
The stability modes are explored by using the Fourier decomposition of the master variable, $\Psi = \int \Psi_{\omega}e^{i\omega t}dt$, we obtain
\begin{eqnarray}
\mathcal{H}\Psi_{\omega} &:=& -\dfrac{\partial^{2}\Psi_{\omega}}{\partial r^{*2}} + \left[\lambda \dfrac{A\mathcal{S}}{C\mathcal{W}}-\dfrac{1}{2\mathcal{S}}\dfrac{\partial^{2}\mathcal{S}}{\partial r^{*2}}+\dfrac{3}{4\mathcal{S}^2}\left(\dfrac{\partial\mathcal{S}}{\partial r^{*}}\right)^2+\dfrac{1}{2C\mathcal{S}}\dfrac{\partial\mathcal{S}}{\partial r^{*}}\dfrac{\partial C}{\partial r^{*}}-\dfrac{1}{2C}\dfrac{\partial^{2}C}{\partial r^{*2}}\nonumber \right. \\ 
&+& \left. \dfrac{3}{4C^2}\left(\dfrac{\partial C}{\partial r^{*}}\right)^2\right]\Psi_{\omega} =\dfrac{\mathcal{S}}{\mathcal{P}\mathcal{W}}\omega^2\Psi_{\omega},\\
&=& -\dfrac{\partial^{2}\Psi_{\omega}}{\partial r^{*2}} + V\Psi_{\omega} = \omega^{2}_{eff}\Psi_{\omega}.   
\label{eq:schr}
\end{eqnarray}
The previous expression is a generalization of the Regge-Wheeler equation \cite{regge}, we must note that the scalar field perturbation vanished but the Eq. (\ref{eq:schr}) depends on the backreaction produced by the scalar field. In general, the spectrum of the operator $\mathcal{H}$ is positive definite always that
\begin{equation}
\int dr^*(\Psi_{\omega})^{*}\mathcal{H}\Psi_{\omega} = \int dr^*\left[\left|D\Psi_\omega \right|^2+V_{\alpha}\left|\Psi_\omega \right|^2\right]- \left. (\Psi_\omega D \Psi_\omega)\right|_{Boundary}, \label{eq:spect}
\end{equation}
where the operator $D$ has the form
\begin{equation}
D=\dfrac{\partial}{\partial r^*}+\alpha,
\label{eq:operator}
\end{equation}
and $V_{\alpha}=V+\dfrac{d\alpha}{dr^*}-\alpha^2$. If we choose $\alpha = \dfrac{1}{2C}\dfrac{d C}{dr^{*}}+\dfrac{1}{2\mathcal{S}}\dfrac{d\mathcal{S}}{dr^{*}}$, we can obtain the Regge-Wheeler potential
\begin{equation}
V_{\alpha} = \lambda \dfrac{A\mathcal{S}}{C\mathcal{W}}. 
\label{eq:potential}
\end{equation}
Always that $V_{\alpha} \geq 0$, we have a stable configuration, in order to satisfy this condition we must have vanishing perturbations at the horizon, i.e., $\left. (\Psi_\omega D \Psi_\omega)\right|_{Boundary} = 0$. In general, near the black hole horizon, $r_{H}$, in order to have vanishing perturbations we could assume an expansion of the following form: $\Psi_{\omega} = \sum^{\infty}_{n=0} \gamma_{n}(r-r_{H})^{n}$, where $\gamma_{n}$ is a constant coefficient, however we must check also the behavior of the operator $D$, since involves derivatives of the scalar field and the function $B$. The action of the operator $D$ over $\Psi_{\omega}$ can be written as follows
\begin{equation}
D\Psi_\omega = \sqrt{\frac{A}{B}}\frac{\partial \Psi_{\omega}}{\partial r}+\frac{1}{2C}\sqrt{\frac{A}{B}}\frac{\partial C}{\partial r}\Psi_{\omega}+\frac{1}{2}\left\lbrace \sqrt{\frac{A}{B}}\frac{1}{\left(1-\dfrac{\phi_{0r,r}}{B}+\dfrac{\phi_{0,r}B_{r}}{2B^2}\right)}\frac{\partial}{\partial r}\left(1-\dfrac{\phi_{0r,r}}{B}+\dfrac{\phi_{0,r}B_{r}}{2B^2}\right)\right\rbrace \Psi_{\omega}, 
\label{eq:horizon}
\end{equation}
where we have used the change, $\dfrac{\partial}{\partial r}=\sqrt{\dfrac{B}{A}}\dfrac{\partial}{\partial r^{*}}$ defined previously together with the Eqs. (\ref{eq:f2}) and (\ref{eq:operator}).\\ 

In terms of the black hole solution obtained in the previous section we have $B = 1/A = 1/f(z)$, therefore $A/B = f^{2}(z)$. In order to visualize the behavior of the perturbations at the horizon, we will consider the solution given in Eq. (\ref{eq:solution2}) and the background field, $\phi_{0}(r)$, will be given by the profile written in Eq. (\ref{eq:scalar}), at the horizon we have $f(z_{H}) = 0$, then the first two terms on the r.h.s. of Eq. (\ref{eq:horizon}) are zero. Now, for the third term we have some derivatives of the scalar field and the function $B$ or $f(z)$, in generic form we can write the third term as follows
\begin{equation}
\left\lbrace \frac{f^{2}(z)\Gamma(z)}{1-f(z)\left[\frac{\sqrt{2} \nu  (1-z)^3}{1+\nu  (1-z)}-\frac{\nu ^2 (1-z)^2}{\sqrt{2} (1 +\nu  (1-z))^2} \right]-\frac{f^{2}(z)}{2}\left[\frac{\nu  (1-z)^2}{\sqrt{2} (1+\nu  (1-z))} \right]f'(z)}\right\rbrace \Psi_{\omega},
\label{eq:third}
\end{equation}
where $\Gamma(z)$ is a cumbersome function involving derivatives of the scalar field and the hypergeometric function, however we can factorize the function $f(z)$ from its terms. See for instance that the first derivative of the function $f(z)$ takes the following form \cite{abramowitz}: $f'(z) = - \beta  z (1-z)^{\beta+1} \, F_1(a,b,2;z)+(1-z)^{\beta+2 } \, F_1(a,b,2;z)+\frac{a b}{2}z (1-z)^{\beta+2 } \, F_1(a+1,b+1,3;z)$ which results $f'(z=z_{H}) = 1$, for the second derivative $f''(z)$ we have a constant value when we evaluate at the horizon. Besides, all the derivatives of the scalar field take a constant value for $z=z_{H}$, therefore the denominator of Eq. (\ref{eq:third}) is equal to one at the horizon and the numerator is zero, then we can see that the perturbations vanish at the horizon. In consequence, only by determining the behavior of the potential, $V_{\alpha}$, we can establish a stability criterion for the black hole solution.\\  

\begin{figure}[htbp!]
\centering
\includegraphics[width=9.5cm,height=6.5cm]{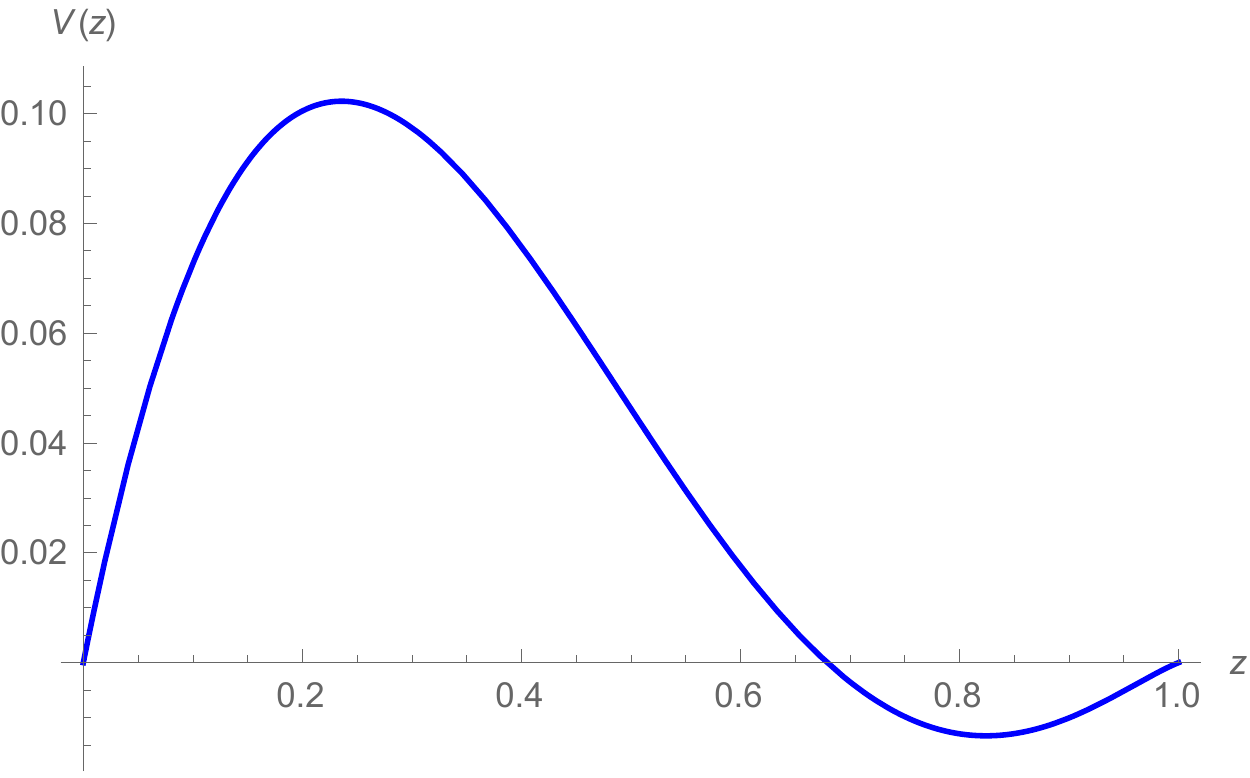}
\includegraphics[width=9.5cm,height=6.5cm]{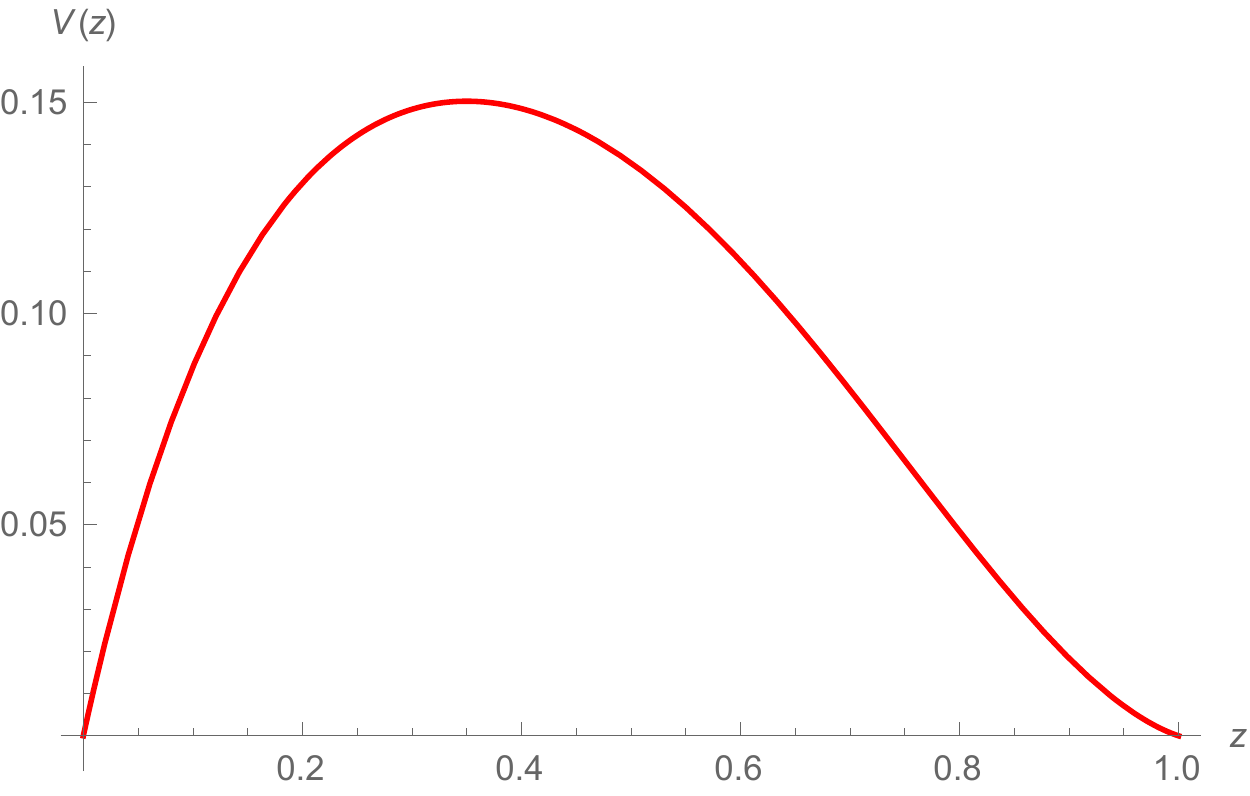}
\caption{Behavior of the $V_{\alpha}$ potential as a function of the coordinate $z$. The upper panel corresponds to $\nu = 0.001$ and the lower panel to $\nu = 0.6$. For the value $\beta = -1$ we obtain a similar shape for the potential as the one shown in both panels.} 
\label{fig:potential}
\end{figure}

In Fig. (\ref{fig:potential}) we depict the behavior of the potential given in Eq. (\ref{eq:potential}), for this purpose we use the expression written in (\ref{eq:solution1}), which corresponds to a solution of the Einstein-Gauss-Bonnet model and depends on the parameter $\nu$ introduced in the profile for the scalar field given in Eq. (\ref{eq:scalar}). Besides the components of the metrics (\ref{eq:ansmetric}) and (\ref{eq:pert}) can be related directly. We have considered the integration constants $A=B=1$ and $\lambda \geq 0$.\\

In general, we will have some variations of the solution (\ref{eq:solution1}) according to the combination of signs of $a_{\mp}$ and $b_{\pm}$, given in (\ref{eq:coeff}). In the Fig. (\ref{fig:potential}), we show the case in which $c \geq 2$ together with $a_{-}$, $b_{+}$ and $\beta=0$, a similar behavior as the one shown in the plots can be obtained for $c \leq 0$. If we choose the cases $a_{-}, b_{-}$ and $a_{+}, b_{+}$, the potential $V_{\alpha}$ is always positive regardless the value of the parameter $\nu$. For the latter case, $a_{+}, b_{+}$, the potential exhibits a growing behavior as we approach to $z=1$. For the election $a_{+}, b_{-}$, the resulting potential is the same of Fig. (\ref{fig:potential}).\\
 
As can be seen in the upper panel of Fig. (\ref{fig:potential}), the potential becomes negative at large distances, $z(r\rightarrow \infty)\rightarrow 1$, therefore the model exhibits an unstable behavior under odd parity perturbations, however, the potential will be always positive if we increase the value of the parameter $\nu$ (as shown in the lower panel), then, always that we choose the appropriate value of $\nu$, the model will be stable under odd parity perturbations.

\section{Final remarks}
\label{sec:final}

In this work we focused on the study of the stability of an Einstein-Gauss-Bonnet hairy black hole solution under odd parity perturbations. The black hole solution under study was obtained by considering a real profile for the scalar field that is regular everywhere, at large distances we have, $\phi \rightarrow 0$, and at short distances the scalar field becomes relevant but exhibits an asymptotic growth.\\

We can visualize the singular nature of the obtained solution despite it depends on the hypergeometric function since we have a linear dependence on the coordinate $z$, being $z=0$ the location of the horizon event and $z=1$ corresponds to infinity. The obtained solution can be written in two different forms according to the following considerations: $c = n \geq 2$ and $c = - n \leq 0$, where $n$ is an integer number and $c$ is a parameter of the hypergeometric function, however in both cases the linear dependence on the coordinate is present. As shown explicitly in this work, for the case $c = n \geq 2$ we can recover the Schwarzschild black hole from our solution.\\

By considering the most general static background solution for a scalar-tensor theory we can observe that as in previous results \cite{anabalon, cisterna}, the equations of motion of the model can be linearized and the contribution of the scalar field to the dynamics is only through the backreaction of the background solution. Then, by constructing the generalized Regge-Wheeler equation we can obtain the stability criterion for the Einstein-Gauss-Bonnet model, which corresponds to the positivity of the potential, $V_{\alpha}$. The shape for the resulting potential can be formulated from the obtained solution for the model. As shown in the work, we have vanishing perturbations at the horizon, then the positivity of the potential guarantees the stability of the black hole under odd parity perturbations. On the other hand, the potential $V_{\alpha}$ depends explicitly on the parameter $\nu$, this parameters enters in the definition of the scalar field profile, for certain values of this parameter the model exhibits instabilities, however the stability of the model will be guaranteed as long as an adequate choice of $\nu$ is made. It is worthy to mention that the shape of the potential is the same for both cases in which the parameter $c$ is different.

\section*{Acknowledgments}
M.C. work has been supported by S.N.I. (CONACyT-M\'exico). G. A. P. and I. M. R. also acknowledge support from CONACyT trough a doctoral and master grants, respectively.

\end{document}